\UseRawInputEncoding

\documentclass[twocolumn]{aastex7}
\usepackage{amsmath}
\usepackage{bm}
\usepackage{gensymb}
\usepackage{caption}
\usepackage{float}

\shorttitle{Crab pulsar}
\shortauthors{Liu et al.}
\graphicspath{{./}{figures/}}
\begin{document}

\title{Evolution of Crab pulsar: magnetic inclination angle and spin
}

\correspondingauthor{Jian-Min Dong}
\email{dongjm07@impcas.ac.cn}

\author{Cong-Xing Liu}
\affiliation{Institute of Modern Physics, Chinese Academy of Sciences, Lanzhou 730000, China}
\affiliation{School of Physical Sciences and Technology, Hebei University, Baoding 071000, China}
\email{lcx@impcas.ac.cn}

\author{Jian-Min Dong}
\affiliation{Institute of Modern Physics, Chinese Academy of Sciences, Lanzhou 730000, China}
\affiliation{School of Nuclear Science and Technology, University of Chinese Academy of Sciences, Beijing 100049, China}
\email{dongjm07@impcas.ac.cn}
%\collaboration{6}{(AAS Journals Data Editors)}



\begin{abstract}

The well-observed Crab pulsar helps one to uncover the underlying knowledge about pulsar evolution. The routine evolution model simultaneously describes the spin-down caused by the magnetic dipole radiation (MDR) and gravitational wave radiation (GWR), damping of the free-body precession owing to the bulk viscosity, and GWR-induced quenching of the magnetic inclination angle $\chi$. We explore the pulsar evolution based on this routine model supplemented with the effects of shear viscosity, r-mode, electromagnetic torque, and accretion, respectively, with the stellar thermal evolution as an important input. The impact of shear viscosity on radio-pulsar evolution is negligible, as it only slightly increases the magnetic inclination angle and promotes spin-down in magnetars. Under the observational limit for its saturation amplitude, the r-mode also turns out to be completely negligible.
Yet, the electromagnetic torque (under certain conditions), along with the accretion based on our three-dimensional fallback disk accretion model, are all shown to suppress the growth of the magnetic inclination angle. 
When applied to the Crab pulsar, the routine model best reproduces the magnetic inclination angle $\chi$, the spin period $P$, and the spin period derivative $\Dot{P}$ simultaneously, indicating the important role of bulk viscosity. The inclusion of the electromagnetic torque and accretion works even worse, suggesting these two factors perhaps are overestimated for Crab pulsar. Intriguingly, the calculated magnetic inclination angle derivative $\Dot{\chi}$ is $(6.3\times 10^{-3} - 0.3)\, {\rm degree/century}$ with the routine model, also in agreement with the observed tiny $\Dot{\chi} = 0.62\, {\rm degree/century}$.

\end{abstract}



\section{Introduction} \label{sec:intro}
The famous Crab pulsar (PSR B0531+21) originated in the collapse of the core of a massive star in 1054 AD, was firstly discovered in 1968 (Lovelace et al. \citeyear{Lovelace_1968Pulsar}; Staelin \& Reifenstein \citeyear{Staelin1968Pulsating}). Since then, this pulsar has been continuously monitored for half a century, yielding fruitful and accurate observational data (Lyne et al. \citeyear{Lyne_1993_Crab}, \citeyear{Lyne_2013_Evolution}, \citeyear{brakingindiceslt3}). These data provide an excellent opportunity to shed light on the pulsar evolution and even the properties of dense matter inside the star. The evolution of pulsars, that is, the magnetic inclination angle $\chi$ (an angle between the magnetic axis and the spin axis), spin frequency $\nu$, and stellar temperature changes over time, has drawn great interest in recent years. It should be emphasized that the evolution of these three physical quantities is coupled together, making the problem actually somewhat complex.

The spin-down of an isolated NS is primarily driven by the loss of spin energy through magnetic dipole radiation (MDR), assuming a vacuum magnetosphere, in other words, neglecting plasma effects (Goldreich \citeyear{Goldreich1970Neutron}). This mechanism follows a power-law relation between the spin frequency and its derivative: $\Dot{\nu} = -K\nu^3$. 
Since the braking index is defined as $n = \nu \Ddot{\nu}/\dot{\nu}^2 = 2-P\Ddot{P}/\dot{P}^2$, predicted to be $n=3$ by this model. However, this braking index is generally inconsistent with the observations (Lyne et al. \citeyear{brakingindiceslt3}; Archibald et al. \citeyear{brakingindicesgt3}), suggesting that additional mechanisms take effect. 
Several mechanisms cause $n < 3$, including braking torques from relativistic particle winds (Xu \& Qiao \citeyear{Xu_2001}; Kou \& Tong \citeyear{Kou2015Rotational}; Gao et al. \citeyear{Gao_2016Constraining}; Tong \& Kou \citeyear{Tong_2017Possible}), accretion of the fallback disk around a NS (Menou et al. \citeyear{Menou_2001_Disk}), increasing in the surface magnetic field due to the magnetic field being buried by accretion and diffused to the surface (Ho \citeyear{Ho2015Magnetic}; Ek{\c{s}}i \citeyear{Yavuz2017magnetic}), etc. And for $n>3$, it could be caused by magnetic dipole spin-down in a plasma-filled magnetosphere (Spitkovsky \citeyear{Spitkovsky_2006_Time}; Philippov et al. \citeyear{Philippov_2014_Time}; Ekşi et al. \citeyear{Ekşi_2016}), gravitational wave radiation (GWR) (de Araujo et al. \citeyear{Araujo_2016}), etc. Accurately describing the braking index does not mean that the spin period $P$, its derivatives $\dot{P}$, and $\Ddot{P}$ can be well reproduced.

Among these, GWR contributes to NS spin-down through two main channels: stellar deformation due to the strong magnetic field (Cutler \& Jones \citeyear{Cutler2000Gravitational}) and r-mode (Andersson et al. \citeyear{Andersson_1998_Class}).
In highly magnetized pulsars, magnetic deformation is the dominant contributor (Cutler \citeyear{Cutler2002Gravitational}). For most pulsars, the quadrupole ellipticity of magnetic deformation is negative, so the toroidal magnetic field is stronger than the surface polar magnetic field. The r-mode is able to facilitate spin-down by GWR theoretically (Owen et al. \citeyear{Owen_1998_gravitational}; S\'a \& Tom\'e \citeyear{Paulo_2005_nonlinear}; Cheng \& Yu \citeyear{Cheng_2014_HOW}). In this work, we re-examine the effects of r-mode under new observational constraints, where its saturation amplitude is limited to less than $5\times10^{-8} (R/12\, \rm{km})^{-2}$ (Haskell \& Patruno \citeyear{Haskell_2017_Gravitational}).

Some new-born NSs may be accompanied by the fallback disk.
In the fallback disk accretion model, the accretion phase increases the angular momentum of the NS, leading to spin-up, while the propeller effect loses angular momentum, causing spin-down. This model predicts a braking index of 2.51 for the Crab pulsar (Menou et al. \citeyear{Menou_2001_Disk}), consistent with observations. However, in a system comprising a fallback disk and an NS, the spin-disc angle $\theta$, defined as the angle between the spin axis of the NS and the disk axis, plays a crucial role. To account for this, we propose a new three-dimensional accretion evolution model in this study.

The evolution of spin frequency $\nu$ and the evolution of magnetic inclination angle $\chi$ are coupled together.
This evolution of $\chi$ occurs in two stages. In the first stage, the temperature inside the star is too high to form  $\rm ^3PF_2$ neutron superfluidity in the stellar core. In the second stage, as the temperature drops below the critical temperature $T_c$, $\rm ^3PF_2$ neutron superfluidity occurs (Dall'Osso et al. \citeyear{Dall'Osso_2009_Early}), which is crucial for the evolution of the magnetic inclination angle (Cheng et al. \citeyear{cheng2018probing}).

During the first stage, the evolution of the magnetic inclination angle is primarily governed by the interplay between the bulk viscosity of stellar matter and GWR. The shear viscosity has little effect on the magnetic inclination angle (Lander \& Jones \citeyear{Lander_2018_neutron}). For a biaxial ellipsoid, bulk viscosity drives the spin and symmetry axes toward alignment (orthogonality) if the shape is oblate (prolate) (Mestel \& Takhar \citeyear{Mestel_1972_internal}), resulting in the magnetic inclination angle decreasing (increasing) accordingly. Observations of the Crab pulsar indicate an increasing magnetic inclination angle, perhaps suggesting that its shape is distorted into a prolate ellipsoid (Lyne et al. \citeyear{Lyne_2013_Evolution}). This implies that the magnetic field is primarily dominated by a toroidal component.  
Furthermore, this distinctive feature that the magnetic inclination angle $\chi$ of the Crab pulsar grows could serve as a key motivation for the inclusion of fallback accretion (Biryukov \& Abolmasov \citeyear{Biryukov_2021_Magnetic}; Yang \& Li \citeyear{Yang_2023_Magnetic}) and internal energy dissipation.
As the temperature decreases, bulk viscosity weakens, while shear viscosity becomes more significant. When the temperature falls below a critical temperature $T_c$, bulk viscosity is significantly suppressed, whereas shear viscosity is only partially reduced (Shang et al. \citeyear{SHANG2020135963}). If the critical temperature $T_c$ is low as predicted by Dong et al. (\citeyear{Dong_2016_role}), the shear viscosity is expected to be important.
The present work focuses on the evolution of the young Crab pulsar and does not involve the second stage, such as the effect of the core-crust coupling owing to the onset of superfluidity (Alpar \& Sauls \citeyear{Alpar_1988_On}).

The structure of this paper is organized as follows. We present the routine pulsar model in Section \ref{Sec: The typical pulsar model}. In section \ref{Sec: Model modification}, we improved the routine model by introducing the shear viscosity, the r-mode, the electromagnetic torque, and the accretion, and then assess their roles in the evolution of spin and of the magnetic inclination angle $\chi$. In section \ref{sec: Crab pulsar}, we attempt to explain the observations of Crab pulsar within these models. Finally, a summary is briefly presented in Section \ref{sec:summary}.

\section{routine model}
\label{Sec: The typical pulsar model}

The magnetic inclination angle $\chi$, the angular frequency $\Omega = 2\pi \nu$, and the temperature $T$ are coupled together.
The dissipation of spin energy in an NS is primarily driven by two mechanisms: GWR and MDR in a vacuum. GWR arises from stellar deformation driven by the toroidal magnetic field.
The evolution of pulsar angular frequency $\Omega$ versus time is expressed as
\begin{flalign}\label{Eq: the evolution of angular frequency}
   &\ \Dot{\Omega} =  - \frac{B^2_{d} R^6\Omega^3}{6Ic^3}\sin^2{\chi}  -\frac{2G\epsilon^2_BI\Omega^5}{5c^5}\sin^2{\chi}(15\sin^2{\chi} +1 ), &
\end{flalign}
where $B_d$ is the surface dipole magnetic field strength at the magnetic pole, and $\epsilon_B = -5\Bar{B}^2_tR^4/(6GM^2)$ represents the quadrupole ellipticity due to magnetic deformation (Dall'Osso et al.  \citeyear{Dall'Osso_2009_Early}). Here, $\Bar{B}_t$ is the average toroidal magnetic field component. The moment of inertia $I$ of the NS is determined through an empirical fitting (Raithel et al. \citeyear{Raithel2016Model}).

The magnetic inclination angle $\chi$ of an NS with a liquid core evolves under the combined effects of internal energy dissipation due to viscous processes and GWR expressed as (Dall'Osso et al. \citeyear{Dall'Osso_2009_Early}; Lander \& Jones \citeyear{Lander_2018_neutron})
\begin{equation}\label{Eq: the evolution of magnetic inclination angle}
\Dot{\chi}=-\frac{2G}{5c^{5}}I\epsilon _{B}^{2}\Omega ^{4}\mathrm{{sin}\chi {%
cos}\chi (15{sin}^{2}\chi +1)+\frac{{cos}\chi }{\tau _{vis}{sin}\chi }}.
\end{equation}
The first term represents the quench of $\chi$ driven by GWR, where radiation torque aligns the spin and magnetic axes. The second term on the right-hand side accounts for the influence of stellar viscosity on the damping of free-body precession, with a characteristic damping timescale given by $\tau_{{\rm vis}} = 2E_{\rm {pre}}/\Dot{E}_{\rm {diss}}$, which drives $\chi$ toward $90^\circ$. The free-body precession energy for a prolate ellipsoid ($ \epsilon_B < 0$) is $ E_{{\rm pre}} \simeq  - \frac{1}{2}I\Omega^2\epsilon_B \cos^2{\chi}$ (Dall’Osso et al. \citeyear{Dall'Osso_2009_Early}). The internal energy dissipation rate is expressed as (Lander \& Jones \citeyear{Lander_2018_neutron})
\begin{flalign}\label{Eq: the dissipation energy}
 &\ \begin{aligned}
  \Dot{E}_{{\rm diss}} = & - \int \left[ \eta \left| \nabla \times \boldsymbol{\delta v}\right|^2 + \frac{4}{3} \eta (\nabla \cdot \boldsymbol{\delta v})^2\right]dV \\
   & - \int \xi(r,t) \left| \nabla \cdot \boldsymbol{\delta v} \right|^2 dV ,
   \end{aligned} &
\end{flalign}
where $\boldsymbol{\delta v}$ represents the velocity perturbation. $\xi(r,t)$ is the space- and time-dependent bulk viscosity (BV) coefficient (Liu et al. \citeyear{Liu_2024_effects}), which is quenched by the nucleon-nucleon short-range correlation (Dong \citeyear{Dong2020rmodeIO}).  The shear viscosity (SV) coefficient $\eta$ is generally negligible because it is significantly smaller than $\xi$. The bulk viscosity is far more efficient in dissipating energy and increases the magnetic inclination angle $\chi$ (Lander \& Jones \citeyear{Lander_2018_neutron}).

The viscosity coefficient of the NSs matter sensitively depends on the temperature (Haensel et al. \citeyear{haensel2000bulk}). In this study, the stellar internal temperature $T(r,t)$ is governed by the NS cooling model (Page et al. \citeyear{page2006cooling}):
\begin{flalign}\label{Eq: the cooling}
    &\ \begin{aligned}
        &\frac{d(Le^{2\phi})}{dr} = - \frac{4\pi r^2 e^\phi}{\sqrt{1 - 2Gm/c^2r}}(C_V\frac{dT}{dt} + e^\phi Q_v), \\
        &\frac{d(Te^\phi )}{dr} = - \frac{1}{\kappa}\frac{Le^\phi }{4\pi r^2 \sqrt{1-2Gm/c^2 r}},
    \end{aligned}&
\end{flalign}
where $\kappa$ is the thermal conductivity, and $\phi $ represents the gravitational redshift. Since the stellar interior is not isothermal in the early stages (Sales et al. \citeyear{Sales_2020_Revisiting}), both luminosity $L$ and temperature $T$ vary with time and space. The term $Q_v$ accounts for neutrino emissivity from various cooling processes, which have been extensively studied (Yakovlev et al. \citeyear{yakovlev2001neutrino}, \citeyear{yakovlev1999cooling}). The specific heat capacity $C_V$ is the sum of contributions from both leptons and nucleons (Page et al. \citeyear{page2004minimal}). Furthermore, Dong et al. (\citeyear{Dong_2016_role}) found that nucleon-nucleon short-range correlations suppress neutrino emissivity and nucleon-specific heat capacity, thereby slowing NS cooling. The model mentioned above does not take into account the shear viscosity, r-mode, electromagnetic torque, and accretion. Here we call it the routine model in the following for the sake of discussion.

The NS interior is assumed to consist of $npe\mu$ dense matter without exotic degrees of freedom. The stellar structure of a non-rotating NS is determined by solving the Tolman-Oppenheimer-Volkov equation using an equation of state derived from relativistic mean-field theory. We employ the FSUGarnet interaction (Chen \& Piekarewicz \citeyear{CHEN2015284}), which provides a reliable description of finite nuclei, nuclear matter, and NS structure (Chen \& Piekarewicz \citeyear{CHEN2015284}; Fattoyev et al. \citeyear{fattoyev2020gw190814}). In particular, FSUGarnet predicts a maximum NS mass of $2.07 \pm 0.02 \  \rm M_{\odot}$ (Chen \& Piekarewicz \citeyear{CHEN2015284}), which is consistent with observational constraints on the maximum NS mass (Fonseca et al. \citeyear{Fonseca_2021}).

\section{Model improvement}
\label{Sec: Model modification}

\begin{figure*}
    \centering
    \includegraphics[width=0.9\linewidth]{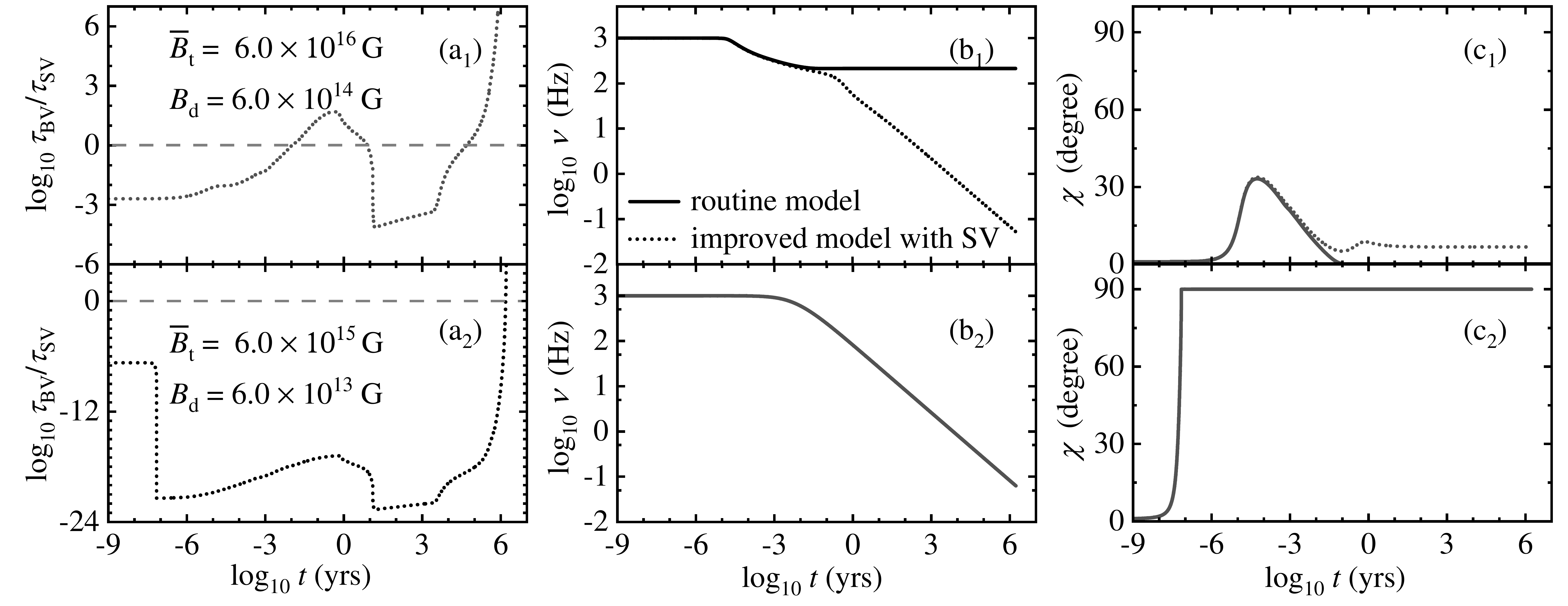}
    \caption{The shear viscosity (SV) effect on the evolution of the damping timescale ratio $\tau_{\rm BV}/\tau_{\rm SV}$, spin frequency $\nu$, and magnetic inclination angle $\chi$ under different magnetic filed. ($\rm a_1 - c_1$) with strong magnetic field: $\Bar{B}_t = 6.0 \times 10^{16} \, \rm{G}$ and $B_d = 6.0 \times 10^{14} \, \rm{G}$; ($\rm a_2 - c_2$) with weak magnetic field: $\Bar{B}_t = 6.0 \times 10^{15} \, \rm{G}$ and $B_d = 6.0 \times 10^{13} \, \rm{G}$.}
    \label{fig: SV_effect}
\end{figure*}

The routine model discussed above overlooks several critical factors. The contribution of shear viscosity to the energy dissipation rate is weak compared to bulk viscosity, but the contribution of shear viscosity increases as the temperature decreases. Additionally, studies have demonstrated that the GWR from the r-mode can influence spin-down processes (S\'a \& Tom\'e \citeyear{Paulo_2005_nonlinear}; Yu et al. \citeyear{Yu2009Long}; Chugunov \citeyear{Chugunov2018evolution}). Moreover, the torque imposed on the star due to the MDR affects the evolution of the magnetic inclination angle $\chi$ (Michel \& Goldwire \citeyear{Michel_190_Alignment}; Lander \& Jones \citeyear{Lander_2018_neutron}). Furthermore, the Crab pulsar, formed by core collapse, may be accompanied by accretion disks formed from incompletely ejected material. Before analyzing the evolution of the Crab pulsar, we firstly improve the routine model by introducing some underlying mechanisms to pulsar evolution: shear viscosity, r-modes, electromagnetic torque, and accretion processes.
To this end, in this section we adopt typical parameter values (not for Crab pulsar specially): a surface dipole magnetic field of $B_d = 5.8 \times 10^{14} \, \rm{G}$ (except in Figure \ref{fig: SV_effect}), a toroidal magnetic field of  $\Bar{B}_t = 6.0 \times 10^{16} \, \rm{G}$ (except in Figure \ref{fig: SV_effect}), a NS mass of  $M = 1.4 \, \rm{M_\odot}$, an initial magnetic inclination angle of  $\chi_i = 1^\circ$, and an initial spin period of  $P_i = 1 \, \rm{ms}$. The initial central temperature is set to  $T(r=0, t=0) = 10^{10} \, \rm{K}$.

\subsection{Shear viscosity}
\label{Sec: the Shear viscosity}

Shear viscosity (SV) of dense matter inside a NS arises from particle-particle scattering, where the primary contributions come from electrons, muons, and neutrons, expressed as  $\eta = \eta_e + \eta_\mu + \eta_n $. The viscosities  $\eta_e$  and  $\eta_\mu$  mainly originate from electron-electron, muon-muon, and electron-muon collisions, as well as interactions with protons via electromagnetic forces. Meanwhile,  $\eta_n $ is determined by neutron-neutron and neutron-proton collisions mediated by strong interactions.

In a young NS, before neutrons or protons enter the superfluid phases, shear viscosity is ineffective in increasing the magnetic inclination angle compared to bulk viscosity (Lander \& Jones \citeyear{Lander_2018_neutron}). However, once  $\rm ^3PF_2$ neutron superfluidity appears in the stellar core, both bulk viscosity and shear viscosity from nucleons (neutrons and protons) are significantly quenched (Shang et al. \citeyear{SHANG2020135963}). Yet, the lepton-lepton collisions remain unaffected (Andersson et al. \citeyear{ANDERSSON2005212}).
At low temperatures, lepton shear viscosity $ \eta_l $ generally exceeds neutron shear viscosity $ \eta_n $. Therefore, the role of shear viscosity in magnetic inclination angle evolution warrants further investigation. In particular, the temperature drops below the critical temperature.

Taking into account the Landau damping in collisions of electrons and muons with charged particles via the exchange of transverse plasmons, the shear viscosity ($l = e $ or $\mu $) is given by (Shternin \& Yakovlev \citeyear{Shternin_2008_shear})
\begin{equation}\label{Eq: the lepton shear viscosity}
\eta _{l}(r,t)=\frac{12\pi c^{2}\hbar ^{3}}{5\lambda }\frac{n_{l}^{2}}{%
q_{t}(\hbar cq_{t})^{1/3}}(k_{B}T(r,t))^{-5/3},
\end{equation}
instead of the standard Fermi-liquid dependence $\eta \propto T^{-2}$. Here $\lambda \approx 6.93$, $k_B$ and $n_l$ are the  Boltzmann constant and number density of lepton $l$.
The $q_l$ is defined as $q_t = \sqrt{\frac{4\alpha}{\pi}\sum_{i} \frac{p_{Fi}^2}{\hbar^2}}$, where $\alpha \approx  1/137$ is the fine structure constant. The shear viscosity of the neutron is clarified based on the Abrikosov and Khalatnikov formulism (Abrikosov \& Khalatnikov \citeyear{Abrikosov_1959_the}; Anderson et al. \citeyear{Anderson_1987_transport}).  The shear viscosity is also space- and time-dependent, similar to bulk viscosity (Liu et al. \citeyear{Liu_2024_effects}), because the density and temperature rely on space and time. Moreover, the effect of the nucleon-nucleon short-range correlation on the shear viscosity is taken into account in this work, which is presented in the explicit discussion in Shang et al. (\citeyear{SHANG2020135963}).

To facilitate cumbersome numerical calculations, we simplify the Equation \ref{Eq: the dissipation energy}. Considering the main contributing terms in the velocity perturbation $\boldsymbol{\delta v}$ in this work leads to $\left| \nabla \times \boldsymbol{\delta v}\right| \approx 15.4 \left| \nabla \cdot \boldsymbol{\delta v}\right|$ (Lander \& Jones \citeyear{Lander_2017_non-rigid}; Lander \& Jones \citeyear{Lander_2018_neutron}). Equation \ref{Eq: the evolution of magnetic inclination angle} for magnetic inclination angle $\chi$ evolution is written as
\begin{eqnarray}\label{Eq: add the SV}
\Dot{\chi} &=&-\frac{2G}{5c^{5}}I\epsilon _{B}^{2}\Omega ^{4}\mathrm{{sin}%
\chi {cos}\chi (15{sin}^{2}\chi +1)+\frac{{cos}\chi }{\tau _{BV}{sin}\chi }}
\nonumber \\
&&\mathrm{+}\frac{\mathrm{{cos}\chi }}{\tau _{\mathrm{SV}}\mathrm{{sin}\chi }%
}.
\end{eqnarray}
$\tau_{\rm{SV}}$ is the dissipation timescale of the shear viscosity and is given by
\begin{flalign}\label{Eq: the shear viscosity dissipation timescale}
    &\ \tau_{\rm{SV}} = \frac{I\Omega^2\epsilon_B \cos^2{\chi}}{238.49 \int\eta(r,t) \left| \nabla \cdot \boldsymbol{\delta v}\right|dV}.&
\end{flalign}
Finally, the results of the coupled Equations \ref{Eq: the evolution of angular frequency}, \ref{Eq: the cooling}, and \ref{Eq: add the SV} elucidate the role of shear viscosity in NS evolution under the strong magnetic field ($\Bar{B}_t = 6.0 \times 10^{16} \, \rm{G}$ and $B_d = 6.0 \times 10^{14} \, \rm{G}$) and the weak magnetic field ($\Bar{B}_t = 6.0 \times 10^{15} \, \rm{G}$ and $B_d = 6.0 \times 10^{13} \, \rm{G}$) respectively, as shown in Figure \ref{fig: SV_effect}.

In the case of the strong magnetic field, as shown in Figures \ref{fig: SV_effect}($\rm a_1-c_1$): the modified Urca processes serve as the primary internal cooling mechanism in a canonical NS with a mass of $1.4\, \rm{M_\odot}$. Their high cooling efficiency, compared to photon emission, results in a rapid temperature drop at the beginning. Furthermore, since $\xi \propto T^6$ and $\eta \propto T^{-5/3}$ or $T^{-2}$, the contribution of shear viscosity to internal energy dissipation increases as the NS cools, as shown in Figure \ref{fig: SV_effect} ($\rm a_1$). After $4$ days, $\tau_{\rm BV}/\tau_{\rm SV} > 1$, shear viscosity dominates internal viscous dissipation. Therefore, Figure \ref{fig: SV_effect} ($\rm c_1$) shows that shear viscosity (dash line) causes a slight increase in the magnetic inclination angle compared to the case of considering only bulk viscosity (solid line) after 4 days. The magnetic inclination angle $\chi$ calculated by the model considering shear viscosity does not evolve to $0\degree$, and thus the MDR and GWR are not fully suppressed, leading to a continued decrease in spin frequency $\nu$ as in Figure \ref{fig: SV_effect} ($\rm b_1$). The case ($\tau_{\rm BV}/\tau_{\rm SV} > 1$) does not persist very long. Because $\tau_{\rm{BV}} \propto \nu^{-2}$ and $\tau_{\rm{SV}} \propto \nu^{-4}$ combined with the rapid decrease in spin frequency after 230 days, the damping timescale due to shear viscosity increases significantly, compared to that due to bulk viscosity, resulting in rapid drops of the ratio $\tau_{\rm BV}/\tau_{\rm SV}$ in Figure \ref{fig: SV_effect} ($\rm a_1$). After 10 years, the dissipation timescale ratio appears to increase slowly. This is due to a decrease in the spin-down rate, which causes the temperature to dominate this ratio. Once the core temperature falls below the $\rm  ^3PF_2 $ neutron superfluidity critical temperature, neutrons become superfluid, leading to suppression of neutron-neutron scattering and a drastic reduction in modified Urca processes due to the superfluid gap. At this stage, electron scattering contributes dominantly to the shear viscosity. As shown in Figure \ref{fig: SV_effect} ($\rm a_1$), the damping timescale ratio increases significantly.

The results of the weak magnetic field case are shown in Figures \ref{fig: SV_effect} ($\rm a_2 - c_2$). Since the magnetic field is weak, the magnetic inclination angle $\chi$ quickly reaches $90\degree$ as in Figure \ref{fig: SV_effect} ($\rm c_2$). This further leads to a rapidly falling of the $\tau_{\rm BV}/\tau_{\rm SV}$ ratio in $\sim 10^{-7}$ years as in Figure \ref{fig: SV_effect} ($\rm a_2$). Moreover, $\tau_{\rm BV}/\tau_{\rm SV} \ll 1$, this indicates that the shear viscosity is insignificant compared to the bulk viscosity. Finally, the spin frequency $\nu$ evolution curve (and the magnetic inclination angle $\chi$ evolution curve) calculated by considering the role of shear viscosity overlaps completely with the one calculated by a routine model in the absent of shear viscosity, as shown in Figures \ref{fig: SV_effect} ($\rm b_2, c_2$).

In a word, in the case of strong magnetic fields, shear viscosity can slightly raise the magnetic inclination angle and promote the loss of spin energy. However, for the Crab pulsar whose surface magnetic field is just $\sim 10^{12}\, \rm G$, the shear viscosity does not affect the evolution of the pulsar as discussed above and hence we can drop it. Therefore, we do not discuss the shear viscosity in Section \ref{sec: Crab pulsar}.

\subsection{R-mode}
\label{sec: the r-mode}

GWR is an important pathway for spin energy loss in newly formed, rapidly rotating NSs. In addition to magnetic deformation, the r-mode is also expected to contribute to this radiation. R-mode is a class of fluid oscillations restored by the Coriolis force, analogous to Rossby waves on the Earth. Due to the Chandrasekhar–Friedmann–Schutz instability, this mode generates gravitational waves in hot, fast-rotating NSs. The emitted waves, in turn, amplify r-mode in the stellar core, increasing their oscillation amplitude and creating a positive feedback loop.
As a result, a part of stellar angular momentum is lost in the form of GWR from the r-mode oscillations theoretically.

Owen et al. (\citeyear{Owen_1998_gravitational}) first presented the model for the evolution of the r-mode due to linear effects, demonstrating that the predominant component of the r-mode is $l=m=2$. However, this does not automatically give the saturation amplitude $\alpha_{\rm{sat}}$ of the r-mode. The saturation amplitude is determined by some non-linear effects. S\'a \& Tom\'e (\citeyear{Paulo_2005_nonlinear}) modified this model, which automatically gave the saturation amplitude of the r-mode by adding some nonlinear effects based on the study of Owen et al. Thus, after accounting for nonlinear effects, the angular momentum of the r-mode is given by $J_c = \alpha^2\Omega(4K + 5)\int_0^R\rho r^6 dr/(2R^2)$, where $K$ is a constant, giving the initial amount of differential rotation associated to the r-mode and determining the saturation amplitude. The range of $K$ is determined by the following conditions: $J_c > 0$, and $\left| J_c \right| \ll I\Omega$. In the case of $J_c > 0 $, this implies that $K > -5/4$ (S\'a \& Tom\'e \citeyear{Paulo_2005_nonlinear}). In the other case $\left| J_c \right| \ll I\Omega$, this implies $K \ll IR^2/(2\alpha^2\int_0^R\rho r^6 dr)$. Consequently, $-5/4<K \ll \alpha^{-2}$ was used in this work.

After considering the r-mode, the conservation of the angular momentum of the star is given by
\begin{equation}\label{Eq: conservation of the angular momentum}
\Dot{J}=-\frac{2J_{c}}{\tau _{\mathrm{GWR(r-mode)}}}-\frac{2J_{c}}{\tau _{%
\mathrm{vis}}}-\frac{2I\Omega }{\tau _{\mathrm{MDR}}}-\frac{2I\Omega }{\tau
_{\mathrm{{GWR(\epsilon _{{B}})}}}},
\end{equation}
where $\tau_{\rm{GWR(r-mode)}}$, and $\tau_{\rm{vis}} = (\tau_{\rm{SV(r-mode)}}^{-1} + \tau_{\rm{BV(r-mode)}}^{-1})^{-1}$ are the r-mode timescale of the GWR from r-mode, and the r-mode timescale of the viscosity, which the explicit discussion are presented in Owen et al. (\citeyear{Owen_1998_gravitational}). $\tau_{\rm{MDR}}$, and $\tau_{\rm{GWR(\epsilon_B)}}$ are the timescale of the MDR, and the GWR from magnetic deformation. Therefore, the evolution of the amplitude of the r-mode $\alpha(t)$ and of the angular frequency of the star $\Omega(t)$ are given by
\begin{flalign}\label{Eq: the evolution of the r-mode}
    \nonumber
    &\ \begin{aligned}
        \Dot{\alpha}=&\ \frac{B_{d}^{2}R^6\Omega ^2\alpha}{12Ic^3}\sin ^2\chi +\frac{G\epsilon _{B}^{2}I\Omega ^4\alpha}{5c^5}\sin ^2\chi \left( 15\sin ^2\chi +1 \right)\\
        & -\left[ 1+\frac{1}{3}\left( 4K+5 \right) Q\alpha ^2 \right] \frac{\alpha}{\tau _{\text{vis}}}\\
	&  -\left[ 1+\frac{4}{3}\left( K+2 \right) Q\alpha ^2 \right] \frac{\alpha}{\tau _{\text{GWR}\left( r-mode \right)}},\\
       \end{aligned}& \\
    &  \begin{aligned}
        \Dot{\Omega}=&\ -\frac{B_{d}^{2}R^6\Omega ^3}{6Ic^3}\sin ^2\chi -\frac{2G\epsilon _{\text{B}}^{2}I\Omega ^5}{5c^5}\sin ^2\chi \left( 15\sin ^2\chi +1 \right)\\
	& + \frac{8}{3}\left( K+2 \right) Q\frac{\Omega \alpha ^2}{\tau _{\text{GWR}\left( r-mode \right)}}+\frac{2}{3}\left( 4K+5 \right) Q\frac{\Omega \alpha ^2}{\tau _{\text{vis}}},\\
    \end{aligned} &
\end{flalign}
where $Q = 3\int_0^R \rho r^6 dr/(2IR^2)$.

To assess the effect of the r-mode on the evolution of NSs, we solve the above coupled equations in the entire parameters space (all possible initial values).
Importantly, Haskell \& Patruno (\citeyear{Haskell_2017_Gravitational}) show that observations of PSR J1023+0038 give an upper limit on the r-mode saturation amplitude, $\alpha_{\rm{sat}} < 5\times 10^{-8} (R/12 \, \rm{km})^{-2}$.
Consequently, when the saturation amplitude is below $4.5 \times 10^{-8}$, r-modes have a minor impact on spin-down only for $K > 10^{14}$.
For example, when the initial values are taken these: $\alpha_i = 1.5\times10^{-50}$, $K = 7.3 \times10^{14}$, $\chi_i = 40.8\degree$, $P_i = 7.8\, \rm ms$, $\Bar{B}_t = 7.6\times10^{16} \, \rm G$, $B_d = 3.4\times10^{12} \, \rm G$, compared to the routine model, after considering the effect of r-mode the period decreased by $7\, \rm ms$ on average. However, its saturation amplitude reaches $2.9 \times 10^{-8}$.
This indicates that the angular momentum associated with r-modes can be comparable to the stellar angular momentum, which is highly unrealistic.
For the evolution of the magnetic inclination angle, only if the initial r-mode amplitude approaches the saturation amplitude and $K>10^{15}$, does the magnetic inclination angle deviate from the result obtained with the routine model by $\sim 0.5\degree$.
In addition, current LIGO and Virgo data do not detect the r-mode, and observations of the long-lived X-ray plateau in the short gamma-ray burst afterglow suggest that gravitational waves originate from magnetic deformations (Xie et al. \citeyear{Xie2022Identifying}). Rajbhandari et al. (\citeyear{Rajbhandari_2021_First}) did not detect r-mode gravitational waves from the Crab pulsar in their analysis of LIGO O1 and O2 data. All of those suggest a weak r-mode.
Therefore, in the condistion of $\alpha_{\rm sat} <4.5 \times 10^{-8}$, the r-modes have no detectable effect on NS evolution in all parameters and can be dropped.
\begin{figure}
    \centering
    \includegraphics[width=0.9\linewidth]{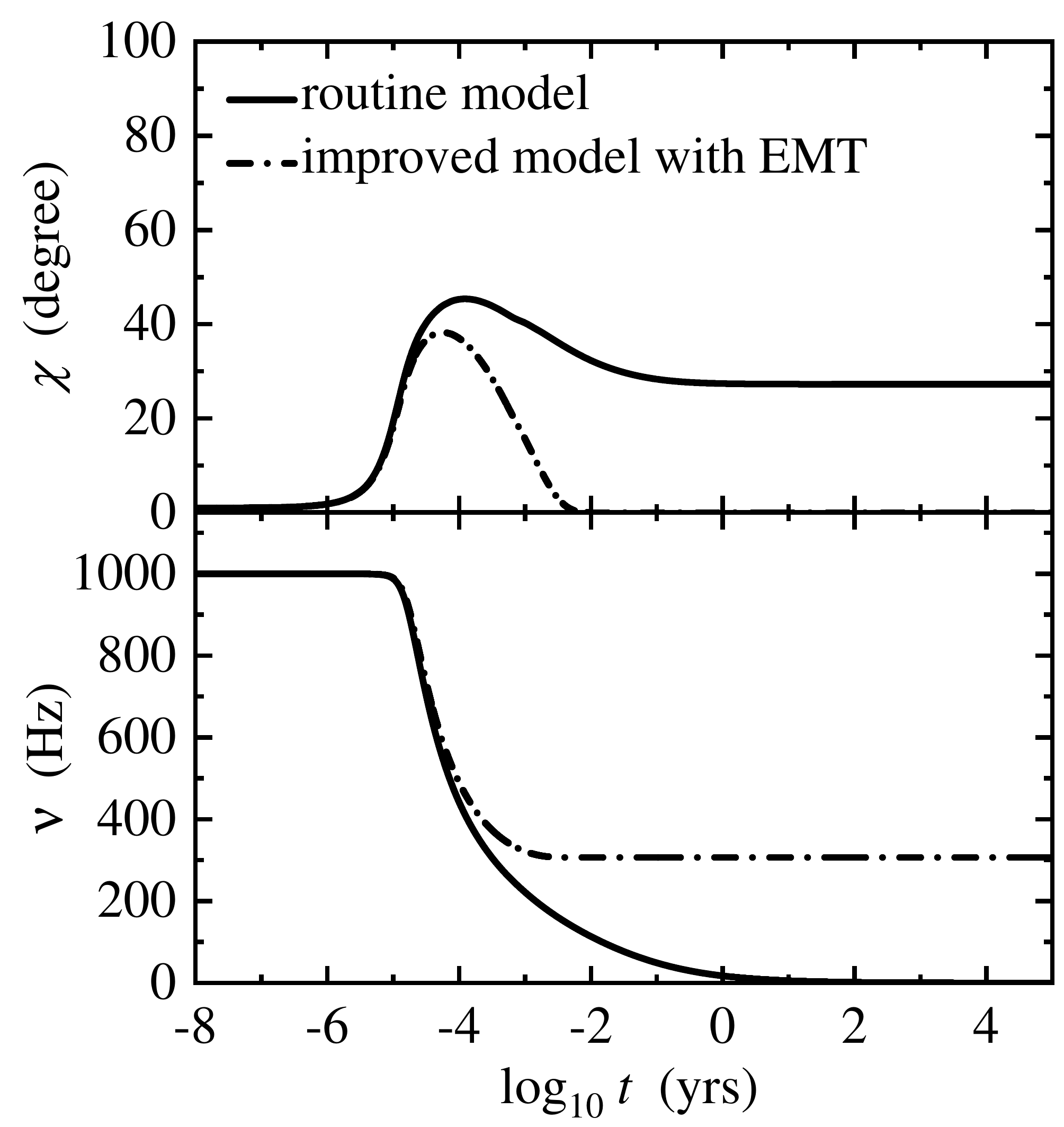}
    \caption{The electromagnetic torque (EMT) effect on the evolution of the spin frequency $\nu$ and of the magnetic inclination angle of a canonical magnetar with $1.4 \, \rm{M_\odot}$.}
    \label{fig:EM_effect}
\end{figure}

\subsection{Electromagnetic torque}
\label{sec: the electromagnetic torque}

A newborn NS is subject to a strong electromagnetic torque. As outlined in Section \ref{Sec: The typical pulsar model}, bulk viscosity dominates the magnetic inclination angle evolution, and GWR torque leads to a decrease in magnetic inclination angle. In addition to these mechanisms, similar to GWR torque, MDR torque also causes the spin axis and magnetic axis to evolve toward an aligned configuration (Michel \& Goldwire \citeyear{Michel_190_Alignment}; Beskin \& Zheltoukhov \citeyear{Beskin2014Anomalous}; Zanazzi \& Lai \citeyear{Zanazzi2015Electromagnetic}). 
If the GWR dominates the magnetic inclination angle $\chi$ decreases, it requires that the ratio of the toroidal field $\Bar{B}_t$ to the dipole field $B_d$ satisfies
\begin{flalign}\label{EQ: MDR and GWR for tilt angle}
    &\ \begin{aligned}
        \Bar{B}_t/B_d > & 1.24 \times 10^2  \left( \frac{M}{1.4\, {\rm M_\odot}}\right)^{1/2} \left( \frac{P}{1\, {\rm ms}}\right)^{1/2} \\
        & \left( \frac{10^{14}\, {\rm G}}{B_d}\right)^{1/2} \left( \frac{12\, {\rm km}}{R}\right)^{3/2}.\\  
    \end{aligned}&    
\end{flalign}
However, a large ratio $\Bar{B}_t/B_d$ would not be expected in a normal pulsar (like the Crab pulsar). Thus, the effect of electromagnetic torque on magnetic inclination angle evolution is not to be ignored.
After introducing the electromagnetic torque, the evolution of the magnetic inclination angle $\chi$ can be written as
\begin{eqnarray}\label{Eq: the effect of the electromagnetic torque}
\Dot{\chi} &=&-\frac{2G}{5c^{5}}I\epsilon _{B}^{2}\Omega ^{4}\mathrm{{sin}%
\chi {cos}\chi (15{sin}^{2}\chi +1)+\frac{{cos}\chi }{\tau _{vis}{sin}\chi }}
\nonumber\\
&&-\frac{B_{d}^{2}R^{6}\Omega ^{2}}{6Ic^{3}}\cos {\chi }\sin {\chi },
\end{eqnarray}
compared with Equation \ref{Eq: the evolution of magnetic inclination angle}.

The electromagnetic torque inhibits the increase of the magnetic inclination angle $\chi$. As shown in Figure \ref{fig:EM_effect}, after $5.0 \times 10^{-5} \, \rm{years}$, it significantly reduces the magnetic inclination angle. After $10^{-2} \, \rm{years}$, the magnetic axis aligns with the spin axis, completely quenching both GWR and MDR, causing the spin frequency to remain constant. 
In addition to the above-mentioned millisecond magnetars with a large ratio $\Bar{B}_t/B_d$, the electromagnetic torque on the magnetic inclination angle evolution is also negligible in pulsars with very weak magnetic fields. This is because bulk viscosity dissipation plays an overwhelming role at this stage, resulting in a quick evolution of the magnetic inclination angle to $90\degree$.

\subsection{Accretion}
\label{sec: the accretion}
For NS, in addition to internal factors influencing their evolution, external factors such as disk accretion may also play a role. A fraction of material from a supernova or merger ejecta may fail to escape and eventually fall back onto the NS. This process occurs when the centrifugal force of the ejected matter is weaker than its gravitational attraction, meaning the inner disk radius $R_{\rm{in}}$ of the accreting material is smaller than the corotation radius $R_{\rm{co}}$. The corresponding accretion torque is given by $T_{\rm{acc}} = \Dot{M}_{\rm{acc}} \sqrt{GMR_{\rm{in}}}$, where $\Dot{M}_{\rm{acc}}$ is the accretion rate.

By balancing the magnetic pressure and the ram pressure of the accretion flow, the inner disk radius can be roughly determined by the Alfve\'n radius as $ R_{\rm{in}} = ( B_d^4R^{12}/GM\Dot{M}_{\rm{disk}}^2)^{1/7}$,
where $\Dot{M}_{\rm{disk}}$ is the mass flow rate of the disk. The proportional coefficient between the inner disk radius and the Alfve\'n radius could range from $0.5$ to $1$ as a result of radiation pressure or some other factor (Ghosh \& Lamb \citeyear{Ghosh_1979_Accretion}; Ghosh \& Lamb  \citeyear{Ghosh_1979_Accretion2}; Arons \& Barnard \citeyear{Arons_1986_Wave}; Arons \& Tavani  \citeyear{Arons_1993_High}). It essentially cannot affect the calculation in this work. Correspondingly, the corotation radius is given by $R_{\rm{co}} = ( GM/\Omega^2 )^{1/3}$.

In other cases, if $R_{\rm{in}} > R_{\rm{co}}$,  the disk material would be accelerated beyond Keplerian velocity, where centrifugal force is stronger than gravitational force. Consequently, centrifugal force ejects this material from the disk, resulting in a propeller outflow. The corresponding torque is given by $T_{\rm{pro}} =  \Dot{M}_{\rm{pro}}R_{\rm{in}}^2\Omega$ where $\dot{M}_{\rm{pro}}$ denotes the mass rate of the propeller outflow (Lai \citeyear{Lai_2014_Theory}). Previous studies assumed that the propeller outflow escapes from the inner radius to infinity once the centrifugal velocity of the disk material exceeds the escape velocity (Illarionov \& Sunyaev \citeyear{Illarionov_1975_Why}; Lovelace et al. \citeyear{Lovelace_1999_Magnetic}; Piro \& Ott \citeyear{Piro_2011_SUPERNOVA}; Metzger et al. \citeyear{Metzger_2018_Effects}).
However, due to the role of viscous dissipation on the accretion disk, the propeller material can not be completely ejected, and eventually fall back into the accretion disk (Li et al. \citeyear{Li_2021_the}; Aloy \& Obergaulinger \citeyear{Aloy_2020_Magnetorotational}).
Therefore, for the propeller-recycling state in this work, we adopt the relations $\Dot{M}_{\rm{rec}} = C_{\rm{rec}} \Dot{M}_{\rm{pro}}$ and $T_{\rm{rec}} = C_{\rm{rec}} T_{\rm{pro}}$ where $C_{\rm{rec}}$ is a damping coefficient representing internal dissipation within the disk and ranges from 0 to 1.

\begin{figure*}
    \centering
    \includegraphics[width=0.9\linewidth]{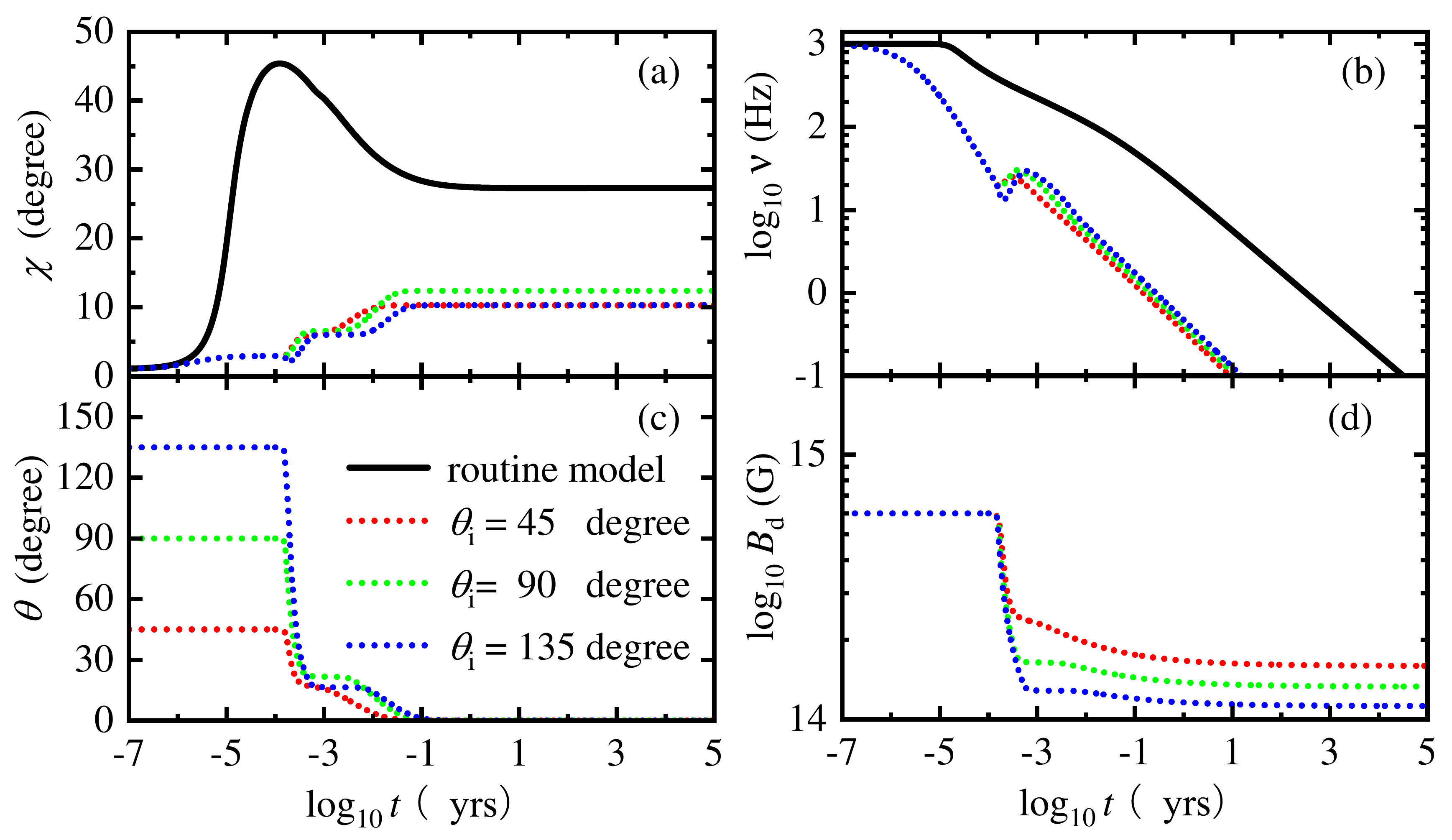}
    \caption{(Color online) The accretion torque effect on the evolution of various physical quantities of a canonical NS with $1.4 \, \rm{M_\odot}$ under different initial spin-disc angle $\theta_i$: (a) the stellar magnetic inclination angle $\chi$; (b) spin frequency $\nu$; (c) the spin-disc angle $\theta$; (d) the stellar surface magnetic field $B_d$.}
    \label{fig: accretion_effect}
\end{figure*}

The evolution of the disk mass and of the disk angular momentum are given by
\begin{flalign}\label{Eq: the evolution of the disk}
    \nonumber
    &\ \begin{aligned}
            \Dot{M}_{\rm disk}= - \Dot{M}_{\rm{acc}} -\Dot{M}_{\rm{pro}} + \Dot{M}_{\rm{rec}},
       \end{aligned}& \\
    & \begin{aligned}
            \Dot{J}_{\rm disk}= - T_{\rm{acc}} -T_{\rm{pro}} + T_{\rm{rec}},
       \end{aligned} &
\end{flalign}
where the disk has been simplified to a ring shape and its mass is concentrated at the outer radius  $R_{\rm{out}}$ (Metzger et al. \citeyear{Metzger_2008_Short}; Lei et al.  \citeyear{Lei_2017_Hyperaccreting}). Then, the angular momentum of the disk can be expressed as $J_{\rm{disk}} \simeq M_{\rm{disk}}\sqrt{GMR_{\rm{out}}}$, where $M_{\rm{disk}}$ is the total mass of the disk. 
The initial disk angular momentum is taken as a typical value $5I\Omega$ (Lei et al. \citeyear{Lei_2017_Hyperaccreting}; Li et al. \citeyear{Li_2021_the}).
Moreover, to describe the continuous transition between the propeller and accretion phases, a parameter $\omega = (R_{\rm{in}}/R_{\rm{co}})^{3/2}$ is defined, with the same procedure as in (Li et al. \citeyear{Li_2021_the}). The accretion rate and the propeller mass loss rate are given by $\Dot{M}_{\rm{acc}} = \Dot{M}_{\rm{disk}}/(1+\omega^n)$ and $\Dot{M}_{\rm{pro}} = \Dot{M}_{\rm{disk}}\omega^n/(1+\omega^n)$ respectively, where the artificial parameter $n > 1$ represents the sharpness of the phase transition, and this does not detract from the discussion. Here the mass flow rate of the disk can be determined by $\Dot{M}_{\rm{disk}} = 0.5 M_{\rm{disk}}\alpha \sqrt{GM/R_{\rm{out}}^3} $, where $\alpha$ is the dimensionless viscosity parameter and this typical value is $0.05$ (Shakura \& Sunyaev \citeyear{Shakura_1973_Black}; Li et al. \citeyear{Li_2021_the}).

For a system composed of a NS accompanied by an accretion disk, previous investigations of the accretion process never considered the effect of the spin-disc inclination angle $\theta$ between the spin axis of the NS and the axis of the disk. However, in a binary system, a model that takes the spin-disc inclination angle into account to trace the NS evolution for both disk-fed and wind-fed NSs is proposed (Biryukov \& Abolmasov \citeyear{Biryukov_2021_Magnetic}; Yang \& Li \citeyear{Yang_2023_Magnetic}). Here, we modify this model to apply to the single NS and disk system.

In this frame, the rotational evolution is given by
\begin{flalign}\label{Eq: the Euler equation}
    &\ \begin{aligned}
            \Dot{\bm{J}}= \bm{N}_{\rm{acc}} + \bm{N}_{\rm{disk-mag}} + \bm{N}_{\rm{psr}} + \bm{N}_{\rm{GWR}}  ,
       \end{aligned}&
\end{flalign}
where $\bm{N}_{\rm{acc}}$, $\bm{N}_{\rm{disk-mag}}$, and $\bm{N}_{\rm{psr}}$ are the accretion torque, the magnetic braking torque caused by disk-magnetic field line interaction, and the pulsar torque, respectively. The explicit discussion is presented in Biryukov \& Abolmasov (\citeyear{Biryukov_2021_Magnetic}). The GWR torque $\bm{N}_{\rm{GWR}}$ caused by magnetic deformation is given by
\begin{flalign}\label{Eq: the gravitational wave radiation torque}
    &\ \begin{aligned}
            \bm{N}_{\rm{GWR}} = \frac{2G}{5c^5}I^2\epsilon_{B}^2\Omega^5\sin{\chi}(15\sin^2{\chi} + 1)\bm{g},
       \end{aligned}&
\end{flalign}
where $\bm{g}$ is the unit vector perpendicular to the magnetic axis.

With the Biryukov \& Abolmasov (\citeyear{Biryukov_2021_Magnetic}) approach, the evolution of an accreting NS can be described by a set of differential equations:
\begin{flalign}\label{Eq: the evolution of an accreting neutron star}
    \nonumber
    &\ \begin{aligned}
           \Dot{\Omega} =&\ - \frac{B^2_{d} R^6\Omega^3}{6Ic^3}\sin^2{\chi} -\frac{2G\epsilon^2_BI\Omega^5}{5c^5}\sin^2{\chi}(15\sin^2{\chi} +1 ) \\
           & + \frac{\Dot{M}_{\rm{acc}}\sqrt{GMR_{\rm{in}}}}{I}\cos{\theta}- \frac{\Dot{I}\Omega}{I} - \frac{B_d^2R^6}{3IR_{\rm{co}}^3},
       \end{aligned}& \\
    \nonumber
     & \begin{aligned}
           \Dot{\theta} = - \frac{\Dot{M}_{\rm{acc}}\sqrt{GMR_{\rm{in}}}}{I\Omega}\sin{\theta} ,
       \end{aligned}& \\
    & \begin{aligned}
           \Dot{\chi} = &\ - \frac{2G}{5c^5}I\epsilon^2_B\Omega^4\rm{sin}\chi\rm{cos}\chi(15\rm{sin}^2\chi + 1) + \frac{\rm{cos}\chi}{\tau_{{\rm vis}} \rm{sin}\chi} \\
           & + \frac{\Dot{M}_{\rm{acc}}\sqrt{GMR_{\rm{in}}}}{I\Omega}  \gamma A(\gamma,\theta,\chi) \sin^2{\theta}\cos{\theta}\sin{\chi}\cos{\chi} \\
           & -\frac{B_d^2R^6\Omega^2}{6Ic^3}\cos{\chi }\sin{\chi }.
       \end{aligned}&
\end{flalign}
$\gamma$ is constant to modulate the accretion torque within the spin period, and we set the typical value $\gamma = 0.99$ as in (Biryukov \& Abolmasov \citeyear{Biryukov_2021_Magnetic}). The expression for $A$ is given by
\begin{flalign}\label{Eq: the A expression}
    &\ \begin{aligned}
            A(\gamma, \theta, \chi) = \left[ 1 - 0.5\gamma(\sin^2{\chi} \sin^2{\theta} + 2\cos^2{\chi} \cos^2{\theta})\right]^{-1} .
       \end{aligned}&
\end{flalign}
The total gravitational mass of the NS can be expressed as $M_{\rm{add-acc}} = M + \int_0^t \Dot{M}_{\rm{acc}} dt'(1 + 3 G \int_0^t \Dot{M}_{\rm{acc}} dt'/(5Rc^2))^{-1}$ (Dai \& Liu \citeyear{Dai_2012_Spin}), leading to a weak change of the moment of inertia $I$.

In addition to the above-mentioned influences on magnetic inclination angle, spin, and stellar mass, the disk also leads to a decrease in the stellar surface magnetic field, as the material falling into the NS surface could repel some open field lines into the closed region and even bury some lines. The corresponding surface magnetic field is determined by the following expression as in (Taam \& Heuvel \citeyear{Taam_1986_Magnetic}; Shibazaki et al. \citeyear{Shibazaki_1989_Does}; Fu \& Li \citeyear{Fu_2013_POPULATION})
\begin{flalign}\label{Eq: the surface magnetic field}
    &\ \begin{aligned}
           B_d (t) = \frac{B_{d,i}}{1 + \int_0^t \Dot{M}_{\rm{acc}} dt' /M_{c}},
       \end{aligned}&
\end{flalign}
where the critical mass $M_c$ is usually in the range $10^{-5} \simeq 10^{-3} \, \rm M_{\odot}$ (Shibazaki et al. \citeyear{Shibazaki_1989_Does}; Zhang \& Kojima \citeyear{Zhang_2006_bottom}). Here we adopt a relatively high value of $M_c = 10^{-3} \, \rm M_{\odot}$.

Figure \ref{fig: accretion_effect} illustrates the impact of accretion on the evolution of NSs. In the initial phase, a relatively low accretion rate results in $R_{\rm{in}} > R_{\rm{co}}$. Consequently, the system enters the propeller phase, in which centrifugal force completely suppresses accretion before $\sim 10^{-4} $ years. As shown in Figure \ref{fig: accretion_effect}(c), the spin-disc angle $\theta$ remains constant when accretion is fully suppressed. Because accretion is inhibited, the surface magnetic field is not buried by infalling material, and therefore, the magnetic field does not decay. However, during this phase, the increase in the magnetic inclination angle is suppressed while the decrease in spin frequency is enhanced, as indicated by the dashed lines in Figures \ref{fig: accretion_effect}(a) and (b). The suppression of the magnetic inclination angle is firstly due to the electromagnetic torque, as discussed in Section \ref{sec: the electromagnetic torque}. Furthermore, the rapid decrease in spin quenches internal energy dissipation, which further limits the growth of the magnetic inclination angle. The enhanced reduction in spin frequency arises from the coupled corotation of the accretion disk with the open magnetic field lines, which effectively drags the NS (Lovelace et al. \citeyear{Lovelace_1995_Spin}).

When $R_{\rm{co}}>R_{\rm{in}}$, the system enters the accretion phase. In this phase, the surface magnetic field decreases sharply as a result of accretion burial, and the spin-disc angle $\theta$ decreases rapidly due to the accretion torque. Both the magnetic inclination angle $\chi$ and the spin frequency $\nu$ increase. However, if the star initially rotates in the opposite direction to the angular momentum of the disc ($\theta_i > 90\degree$), neither the magnetic inclination angle nor the spin frequency immediately increases if the spin-disc angle fails to fall below 90 degrees. In this case, the evolution of the magnetic inclination angle starts with a small angle, which is later compensated by the rapid decay of the magnetic field. This results in the same final value of the magnetic inclination angle whether the initial spin-disc angle is $\theta_i$ or $\theta_i - 90\degree$. Moreover, the closer the initial spin axis parallel to the disk axis, the smaller the magnetic inclination is in the later stages of evolution. When the initial spin-disc angle is larger, it takes a longer time for the spin axis and disc axis to align, leading to the accretion of more material, further burial of magnetic field lines, and a stronger reduction of the magnetic field, as shown in Figure \ref{fig: accretion_effect}(d). In addition, both the magnetic inclination angle and the spin-disc angle exhibit a plateau around $10^{-3}$ years, as shown in Figure \ref{fig: accretion_effect} (a) and (c). This plateau arises because, at that stage, the inner radius $R_{\rm{in}}$ is close to the corotation radius $R_{\rm{co}}$, which suppresses accretion and slows the evolution of both the magnetic inclination angle and the spin-disc angle.

\section{Evolution of Crab pulsar}
\label{sec: Crab pulsar}
With the model and several mechanisms that we discussed above in detail, we now attempt to explain the evolution of the Crab pulsar.
The Crab pulsar is a young radio pulsar with a well-measured period $P_{\rm{Crab}} = 33.08 \, \rm{ms}$, period derivative $\Dot{P}_{\rm{Crab}} = 4.21 \times 10^{-13}\, \rm s/s$, period second-order derivatives $\Ddot{P}_{\rm Crab} = - 2.80 \times 10^{-24} \, \rm s^{-1}$, and hence the braking index $2.51 \pm 0.01$ (Lyne et al. \citeyear{Lyne_1993_Crab}). 
In recent decades, approximately 20 glitches have been observed in the Crab pulsar. However, these events induce tiny changes in spin frequency (typical amplitude $ \Delta \nu/\nu \sim  10^{-9}$ (Lyne et al. \citeyear{brakingindiceslt3}), although in 2017 observation of $ \Delta \nu/\nu \sim  10^{-7}$ (Shaw et al. \citeyear{Shaw2018largest})). Given their relatively small cumulative impact on long-term spin evolution, we neglect the glitch effects in our analysis.
Watters et al. (\citeyear{Watters_2009_ATLAS}) revealed that the magnetic inclination angle of the Crab pulsar is $\chi_{\rm{Crab}} = 55\degree \sim 60 \degree $ (or $70 \degree$) based on computed beaming patterns and light curves from the two-pole caustic emission (or outer-magnetosphere emission) model. Harding et al. (\citeyear{Harding_2008_High}) found that the Crab pulsar profiles and spectrum are well reproduced when the magnetic inclination angle is $45\degree$. Moreover, the separation between the two peaks in the gamma-ray light curve of the Crab suggests a magnetic inclination angle in the range of $60\degree \sim 70 \degree$ (Ardavan \citeyear{Ardavan_2024_Gamma}).
Lyne et al. (\citeyear{Lyne_2013_Evolution}) further measured the rate of change of magnetic inclination angle, $\Dot{\chi}_{\rm Crab} = 0.62\degree \, \rm century^{-1}$.
In this work, we adopt the value of the magnetic inclination angle of the Crab pulsar between 45 and 70 degrees.
Thus, the MDR directly gives the corresponding surface magnetic field strength of $(4.93 \sim  6.55) \times 10^{12} \, \rm{G}$. The actual age of the Crab pulsar is 921 years.

\begin{figure}
    \centering
    \includegraphics[width=1\linewidth]{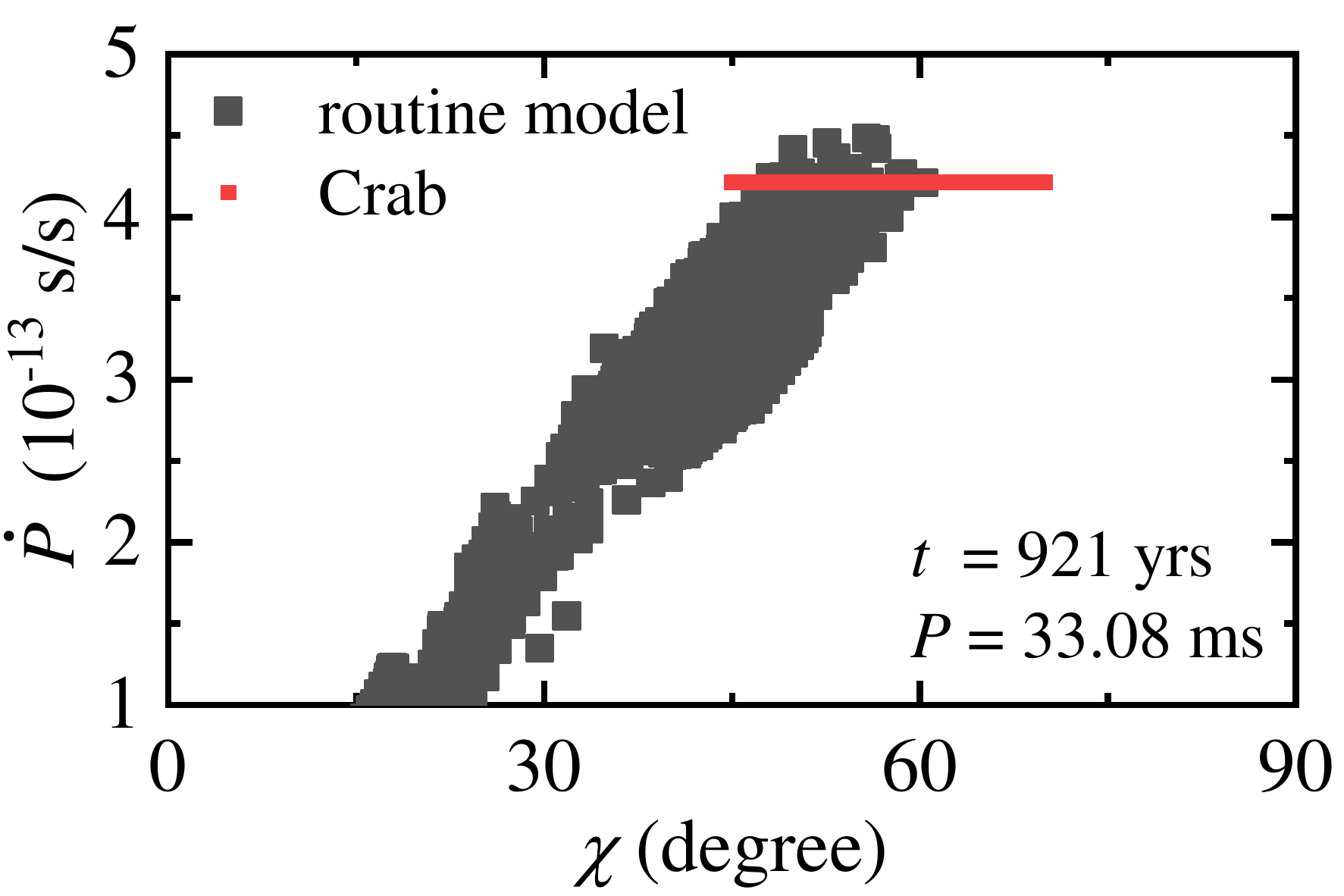}
    \caption{(Color online) Magnetic inclination angle $\chi$ and period derivative $\Dot{P}$ of $1.4\, \rm{M}_\odot$ canonical NSs calculated by the routine model at 921 years, with the initial values of $P_i = (0- 33.08) \, \rm{ms}$, $\chi_i = (0\degree - 70\degree)$,$B_d = (4.93 -  6.55) \times 10^{12} \, \rm{G}$, and $\Bar{B}_t = (B_d - 10^{17}\, \rm G)$. The black region is the calculated results, satisfying the observed Crab pulsar period $P_{\rm Crab} = 33.08\, \rm ms$ at the age of $921 \, \rm yrs$. The red symbol represents the observed $\Dot{P}_{\rm Crab}$ and $\chi_{\rm Crab}$ of the Crab pulsar.}
    \label{fig:Crab_classical}
\end{figure}

\begin{figure}
    \centering
    \includegraphics[width=1\linewidth]{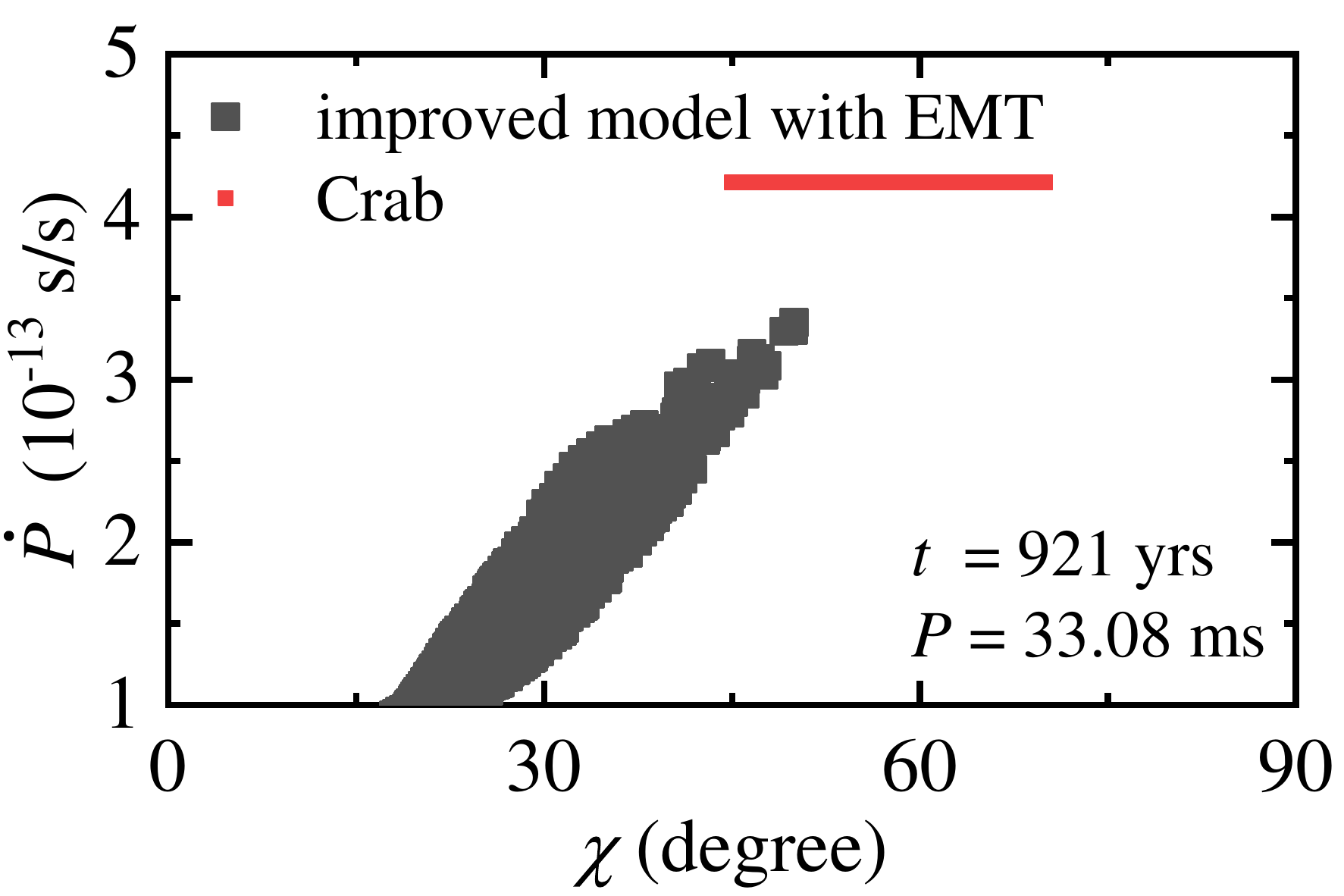}
    \caption{(Color online) The same as Figure \ref{fig:Crab_classical}, but by employing the improved model with the inclusion of the electromagnetic torque (EMT) for magnetic inclination angle.}
    \label{fig:Crab_EM}
\end{figure}

\begin{figure}
    \centering
    \includegraphics[width=1\linewidth]{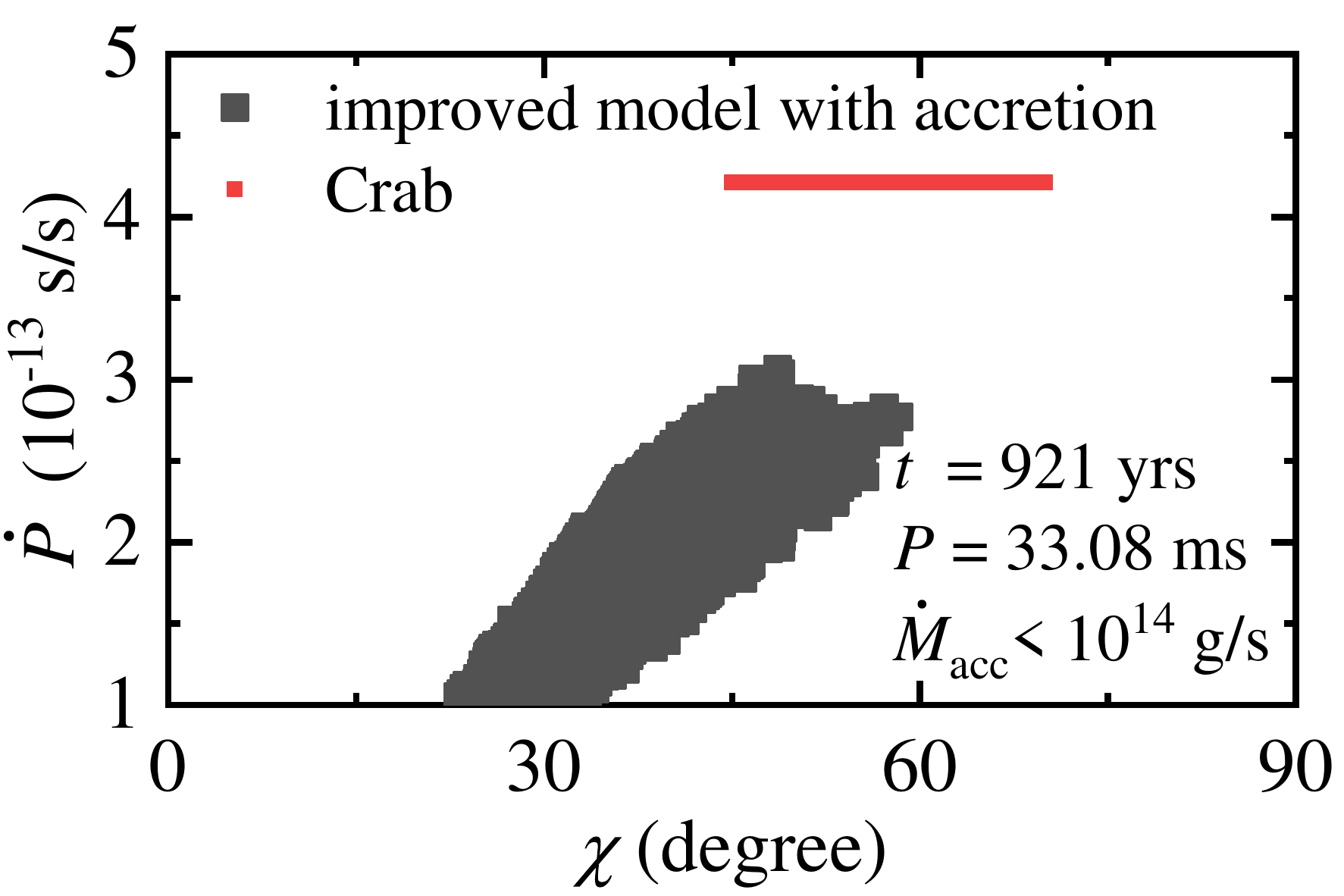}
    \caption{(Color online) The same as Figure \ref{fig:Crab_classical}, but by using the improved model with the accretion correction and accretion rate $< 10^{14} \, \rm g\,s^{-1}$ at the age of $921\, \rm yrs$: $\theta_i = (0\degree - 180\degree)$, $C_{\rm{rec}} = (0- 1)$, $M_{\rm{disk\, i}} = (0 - 0.5)\, \rm{M_\odot}$ (Raskin \& Kasen \citeyear{Raskin_2013_Tidal}; Liu \& Li  \citeyear{Liu_2015_EVOLUTION}), $B_{d\, i} = (10^{12}- 10^{17}) \, \rm{G}$}
    \label{fig:Crab_accretion}
\end{figure}

As discussed in Section \ref{Sec: Model modification}, the influence of the r-mode and shear viscosity on the pulsar rotation and magnetic inclination angle evolution is very weak, and therefore, we do not consider these two factors any longer. Figures (\ref{fig:Crab_classical}-\ref{fig:Crab_accretion}) show the calculated results of the magnetic inclination angle and period derivative with the age of 921 years and the period of $33.08 \, \rm ms$,  under the given initial conditions based on the three models mentioned above: the routine model, the improved model with the electromagnetic torque, and the improved model with the fallback accretion. Since our calculation is performed in an entire parameter space, the final result is a region.

In the routine model, the following entire parameter space (initial values) are adopted: initial period $P_i = (0 - 33.08) \, \rm{ms}$, initial magnetic inclination angle $\chi_i = (0\degree - 70\degree)$, the surface magnetic field $B_d = (4.93 -  6.55) \times 10^{12} \, \rm{G}$, and the toroidal magnetic field $\Bar{B}_t = (B_d - 10^{17} \, \rm G)$. 
With the results of the pulsar evolution calculated via the entire parameter space described above, we obtain the results of the magnetic inclination angle $\chi$ and period derivatives $\Dot{P}$ with the age of 921 years and a period of 33.08 ms, as shown in Figure \ref{fig:Crab_classical}.
The results of the routine model show that the corresponding magnetic inclination angle $\chi$ can range from $46\degree$ to $59\degree$ at 921 years, when the period $P$ is $33.08 \, \rm ms$ and the period derivative $\Dot{P}$ is $4.21 \times 10^{-13}\, \rm s/s$. 
In other words, within a certain range of parameters, the period $P$, the period derivative $\Dot{P}$, and the magnetic inclination angle $\chi$ of the Crab pulsar can be explained simultaneously if the initial conditions are chosen as: $0.08\degree < \chi_i < 0.16\degree$, $15.6 \, {\rm ms} <P_i < 19.0 \, \rm{ms}$ (which is similar to many particle wind acceleration model calculation results that the initial periods are all $\sim 19\, \rm ms$ (Kou et al. \citeyear{Kou2015Rotational})), and $8.3 \times 10^{15} \, {\rm G}<\Bar{B}_t < 1.0\times 10^{16} \, \rm{G}$.
With those values, the magnetic inclination angle derivative $\Dot{\chi}$ is $(6.3\times 10^{-3} - 0.3)\, {\rm degree/century}$, consistent with the value of $\Dot{\chi}_{\rm Crab} = 0.62\, {\rm degree/century}$ (Lyne et al. \citeyear{Lyne_2013_Evolution}). This indicates that the magnetic inclination angle of the Crab pulsar is currently still increasing at a very slow rate, being consistent with observations.
However, with the above initial values, the computed second-order derivatives of the period with respect to time are generally in the range of $ -7.5 \times 10^{-24}\, {\rm s^{-1}}<\Ddot{P} < -4.9\times 10^{-24}\, \rm s^{-1}$, which is larger than the observed value $\Ddot{P}_{\rm Crab} = - 2.80 \times 10^{-24} \, \rm s^{-1}$ in amplitude. The calculated braking index is between 3 and 5.
Since this model only considers MDR and GWR, it can not reproduce accurately the braking index of the Crab pulsars. To explain the braking index, other important factors need to be considered, such as the braking torques from relativistic particle winds (Gao et al. \citeyear{Gao_2016Constraining}; Tong \& Kou \citeyear{Tong_2017Possible}; Zhang et al. \citeyear{Zhang2022Evolution}), magnetic field decay (Gao et al. \citeyear{Gao_2017Dipole}).

Figure \ref{fig:Crab_EM} illustrates the distribution of $\chi$ and $\Dot{P}$ in the improved model with the inclusion of the electromagnetic torque for magnetic inclination angle.  We take the same entire parameters space as in the routine model described above. It is obvious in the profile that theoretically calculated results (black area) deviate from the observed results (red area). Compared with the routine model, by employing the improved model with electromagnetic torque, none of the parameter sets in the entire parameter space can explain the magnetic inclination angle, spin period, and spin period derivative of the Crab pulsar at the same time. The braking of MDR on magnetic inclination angle $\chi$ may be overestimated or the traditional MDR is oversimplified.

Menou et al. (\citeyear{Menou_2001_Disk}) interpreted the braking index of the Crab pulsar by considering accretion as a correction to the typical magnetic dipole model with an accretion rate in the range of $3\times10^{16} \sim 10^{17} \, \rm{g\, s^{-1}}$. The X-ray flow of the Crab pulsar was determined through HXMT by Zhao et al. (\citeyear{Zhao_2023_Long}). In this work, we assume that there is a certain amount of accretion in the Crab pulsar and that the accreted material is all converted to X-rays. Thus, the upper limit on the accretion rate of the Crab pulsar is $10^{14} \, \rm{g\, s^{-1}}$. Moreover, it is assumed that the Crab pulsar magnetic field is buried by accretion. Therefore, the distribution in Figure \ref{fig:Crab_accretion} is a profile qualified by the following two conditions: $\Dot{M}_{\rm acc}^{t = 921\,\rm yrs} < 10^{14}\, \rm{g\, s^{-1}}$, $ 4.93 \times 10^{12} \, {\rm G}< B_{\rm d}^{t = 921\,\rm yrs} < 6.55\times10^{12}\, \rm{G}$. Here, with the addition of a few degrees of freedom due to the consideration of accretion, the entire parameter space is as follows: initial period $P_i = (0 - 33.08) \, \rm{ms}$, initial magnetic inclination angle $\chi_i = (0\degree - 70\degree)$, the initial surface magnetic field $B_{di} = (10^{12}-10^{17})\, \rm{G}$, the initial spin-disc angle $\theta_i = 0\degree - 180\degree$, the initial disk mass $M_{\rm disk\ i} = (0-0.5)\, \rm M_\odot$\footnote{Numerical simulations suggest that white dwarf mergers undergoing accretion-induced collapse (Dessart et al. \citeyear{Dessart2006Multidimensional}, \citeyear{Dessart2007Magnetically}) leave a fossil disk of $\sim (0.05-0.5) \, \rm M_\odot$ around the resulting neutron star. Therefore, here we take the upper limit to be $0.5\, \rm M_\odot$.}, the damping coefficient $C_{\rm rec} = (0-1)$, and the toroidal magnetic field $\Bar{B}_t = (B_d - 10^{17} \, \rm G)$.

The accretion-improved model predicts the braking index range of $n = 2.1\sim 3.9$ for the Crab pulsar when both $\chi$ and $P$ match the observations, accommodating the observed $n = 2.5$. However, as shown in Figure \ref{fig:Crab_accretion}, this model fails to reproduce the spin-period derivatives $\dot{P}$ simultaneously. Specifically, this model, none of the parameter sets in the entire space of parameters, can simultaneously explain the magnetic inclination angle, spin period, spin-period derivatives, and the braking index of the Crab pulsar under the conditions that theoretically calculated accretion rates are lower than the observed maximum possible accretion rate for the Crab pulsar.

Furthermore, we calculate that the interaction time between accretion disk and NS (the time from birth until the inner disk radius $R_{\rm in}$ is greater than the light cylinder radius $R_{\rm lc} = c/\Omega$) is less than 1 year under follows conditions: $\Dot{M}_{\rm acc}< 10^{14}\, \rm{g\, s^{-1}}$ and $P = 33.08\, \rm ms$ at 921 years.
This means that the accretion is perhaps much weaker than estimated, which cannot affect the evolution of the Crab pulsar violently. However, this does not mean that accretion is not important in other NS evolutions. As discussed in Section \ref{sec: the accretion}, strong accretion suppresses the increase in magnetic inclination angle and produces a spin-up phase. Especially in binary systems, accretion is of particular importance in binary evolution (Lai \citeyear{Lai_2014_Theory}; Yang \& Li \citeyear{Yang_2023_Magnetic}; D{\'\i}az Teodori \citeyear{Teodori2025NICER}).

In the above discussion, we utilize the MDR under a vacuum that is widely used to describe spin-down. However, it neglects plasma effects in the pulsar magnetosphere (Goldreich \citeyear{Goldreich1970Neutron}; Faucher-Giguère \& Kaspi \citeyear{Faucher-Giguère_2006}). Philippov et al. (\citeyear{Philippov_2014_Time}) first explained the plasma effects by analyzing the results of time-dependent three-dimensional force-free and magnetohydrodynamic simulations of the pulsar magnetosphere. 

The magnetospheric scenario modifies two key terms in the spin evolution: (1) it replaces the $\sin^2{\chi}$ dependence with $1+\sin^2{\chi}$, and (2) changes the prefactor from $1/6$ to $1/4$ (Spitkovsky \citeyear{Spitkovsky_2006_Time}; Philippov et al. \citeyear{Philippov_2014_Time}). Yet, for the magnetic inclination angle evolution, only the prefactor requires the same modification from $1/6$ to $1/4$.
\begin{figure}
    \centering
    \includegraphics[width=1\linewidth]{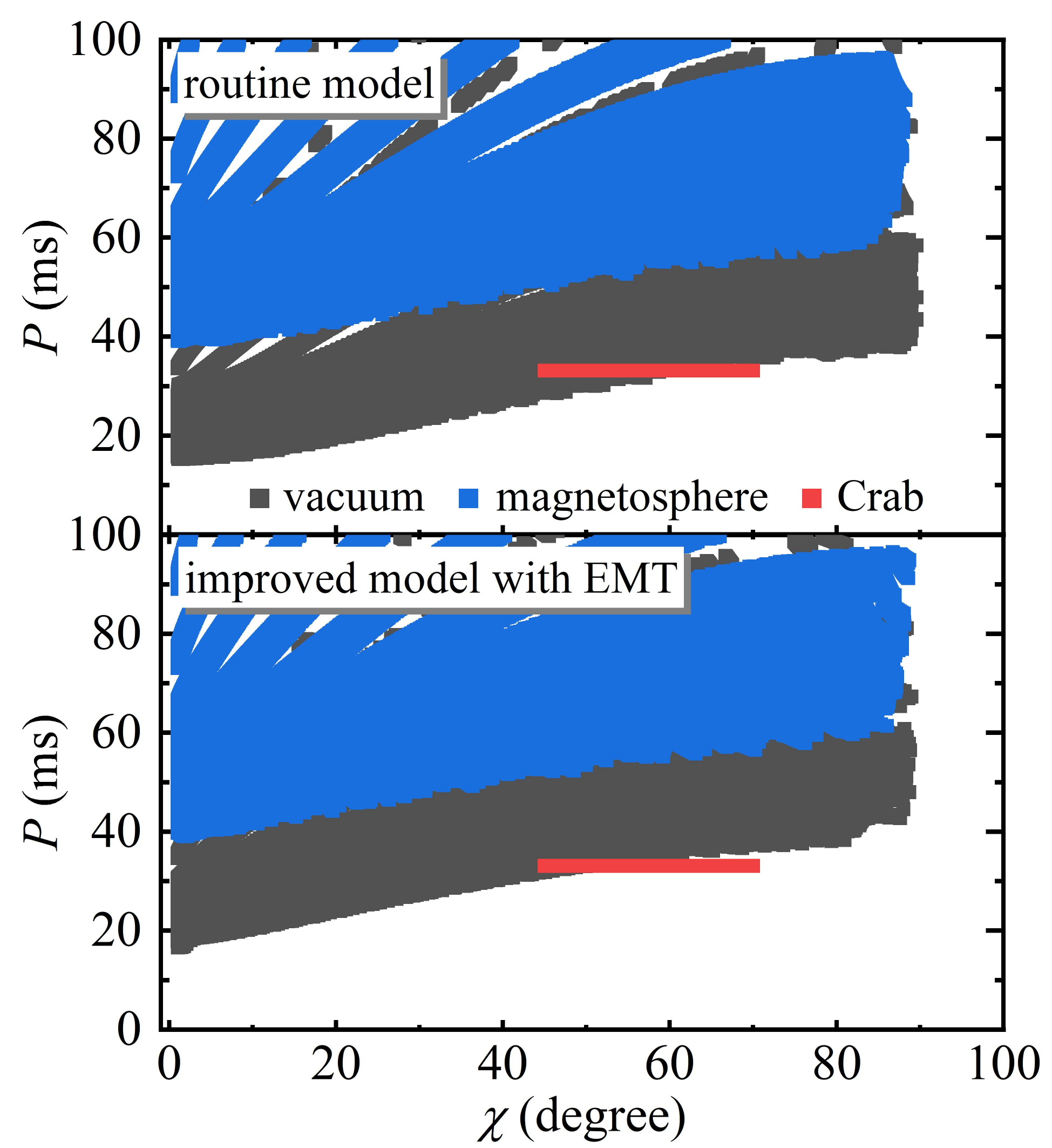}
    \caption{(Color online) The period $P$ and magnetic inclination angle $\chi$ of $1.4 \, \rm M_\odot$ canonical neutron stars calculated under vacuum (black symbol) and in the magnetosphere (blue symbol)  at 921 years, respectively. The upper panel: use the routine model; the lower panel: use the improved model (black symbol) with the inclusion of the electromagnetic torque (EMT) for magnetic inclination angle. The initial values are the same as in Figure \ref{fig:Crab_classical}. The red symbol represents the observed $\chi_{\rm Crab}$ and $P_{\rm Crab}$ of the Crab pulsar.}
    \label{fig:Mag_vac}
\end{figure}
As demonstrated by Equation \ref{Eq: the evolution of angular frequency}, the enhanced spin-down rate in the magnetospheric case leads to more rapid braking compared to the vacuum scenario. Consequently, as shown in Figure \ref{fig:Mag_vac}, the computed periods under magnetospheric conditions are systematically longer than the vacuum case. 
At $t = 921 \, \rm yr$, all magnetospheric models shown in Figure 7 yield periods larger than the observed value of $P_{\rm Crab} = 33.08\, \rm ms$. Furthermore, under the magnetospheric scenario, since the accretion-improved model is analogous to the electromagnetic torque-improved model, it similarly fails to quantitatively reproduce the observed rotation period of the Crab pulsar in this regime, indicating that the MDR under magnetospheric conditions provides a poorer explanation for the observed characteristics of the Crab pulsar.

In summary, through our calculations in the entire parameter space, it is revealed that the routine model without electromagnetic radiation torque and accretion gives the best description of the Crab pulsar evolution.

\section{summary}
\label{sec:summary}

In this study, we focus on the effects of internal shear viscosity, r-mode, electromagnetic torque, and the external accretion disk on NS precession, particularly on the evolution of spin and of magnetic inclination angle. The stellar thermal evolution is an important input to the above evolution. Furthermore, we investigate the evolution of the Crab pulsar and attempt to explain the observed features, including magnetic inclination angle $\chi$, the magnetic inclination angle derivative $\Dot{\chi}$, the period $P$, period derivative $\Dot{P}$, and the period second-order derivative $\Ddot{P}$.

The free-body precession in highly magnetized pulsars involves three key mechanisms: the spin-down caused by the combined effects of MDR and GWR; the increase in the magnetic inclination angle resulting from bulk viscosity damping of precessional motion; and the quenching of the magnetic inclination angle due to GWR. This is described by the routine model which is used in previous investigations.

We modify this routine model by including the space- and time-dependent shear viscosity, r-mode, electromagnetic torque, and accretion, respectively. For the improved model with shear viscosity, the shear viscosity is found to be stronger than the bulk viscosity at the end of the first stage of magnetic inclination angle evolution of magnetars. As a result, shear viscosity slightly elevates the magnetic inclination angle $\chi$ and promotes the loss of spin energy. However, in the case of weak magnetic fields, the contribution of bulk viscosity to the pulsar evolution is drastically larger than that of shear viscosity during the entire evolutionary phase, resulting in a negligible contribution of shear viscosity. Thus, the shear viscosity does not influence the evolution of radio pulsars, such as the Crab pulsars. For the improved model with r-mode, in the limit of the saturation amplitude $4.5\times 10^{-8}$, our calculations show that the r-mode also does not affect the evolution of NSs. 
The electromagnetic torque has a negligible effect on the evolution of the magnetic inclination angle for NSs with either extremely strong (magnetars) or extremely weak magnetic fields.

In the improved model with accretion, we propose a three-dimensional fallback disk accretion model on the basis of the binary star evolution model of Biryukov \& Abolmasov (\citeyear{Biryukov_2021_Magnetic}). Our findings indicate that accretion suppresses the increase of the magnetic inclination angle and that the interaction between the stellar magnetic field and the disk material promotes spin-down. Furthermore, a larger initial spin-disc angle results in more matter accreting onto the NS surface, which promotes the decay of the surface magnetic field. The suppression of the magnetic inclination angle by accretion is more pronounced when the spin axis is nearly parallel to the disk axis.

We have applied the routine model and the improved models to calculate the Crab pulsar evolution. Here, the measurements (including $\chi$, $\Dot{\chi}$, $P$, $\Dot{P}$, and $\Ddot{P}$) of the Crab pulsar allow a more direct test of those models. Under the routine model, we demonstrate that Crab pulsar observations are well reproduced if the initial values are chosen as follows: $0.08\degree < \chi_i < 0.16\degree$, $15.6 \, {\rm ms} <P_i < 19.0 \, \rm{ms}$, and $8.3 \times 10^{15} \, {\rm G}<\Bar{B}_t < 1.0\times 10^{16} \, \rm{G}$. The magnetic inclination angle $\chi$, spin period $P$, and spin-period derivative $\Dot{P}$ of the Crab pulsar are described simultaneously by the routine model. For the magnetic inclination angle derivative $\Dot{\chi}$, our calculation gives $(6.3\times 10^{-3} - 0.3)\, {\rm degree/century}$, which is in agreement with the observed $\Dot{\chi}_{\rm Crab} = 0.62\, {\rm degree/century}$. This suggests that the magnetic inclination angle of the Crab pulsar is currently increasing at a very small rate. The calculated second-order derivative of the spin period gives a result of $ -7.5 \times 10^{-24}\, {\rm s^{-1}}<\Ddot{P} < -4.9\times 10^{-24}\, \rm s^{-1}$, meaning that this routine model is still unable to explain the measured braking index $2.51$ of the Crab pulsar. However, whether the model is improved with electromagnetic torque or accretion, the results calculated in the entire parameter space indicate that, none of the parameter sets in these improved models can explain the magnetic inclination angle $\chi$, the spin period $P$, and the spin period derivative $\Dot{P}$ of the Crab pulsar at the same time. 
For the Crab pulsar, accretion appears too weak to affect its evolution, and the electromagnetic torque model is likely oversimplified, requiring other mechanisms to balance it. Furthermore, within the magnetospheric scenario, all models struggle more to accurately describe its evolution.

Although the routine model can reproduce the magnetic inclination angle, period, and period derivative of the Crab pulsar simultaneously, it still fails to well account for the braking index. The improved models, which take electromagnetic torque and accretion, face even greater difficulty in reproducing these observations.
Perhaps other mechanisms contribute to the evolution, such as the dipole magnetic field decay due to Hall drift and Ohmic dissipation (Jones \citeyear{Jones_1988_Neutron}; Gao et al. \citeyear{Gao_2017Dipole}), the onset of the magnetic multipole field, and wind braking for the young pulsar. 
In addition we do not consider the glitch of the Crab pulsar in this work.
Accordingly, the more comprehensive evolution model of pulsars needs to be further explored.

\begin{acknowledgments}
This work is supported by the National Natural Science Foundation of China under Grants No. 12222511, by the Chinese Academy of Sciences Project for Young Scientists in Basic Research YSBR-088, by the Strategic Priority Research Program of Chinese Academy of Sciences, Grant No. XDB34000000, and by the Continuous Basic Scientific Research Project under Grants No. WDJC-2019-13.
\end{acknowledgments}

\bibliography{References}{}
\bibliographystyle{aasjournalv7}

\end{document}